\documentclass{article}
\usepackage{graphicx}
\usepackage{amssymb}
\usepackage{amsmath}
\usepackage{color}
\newcommand{\ssa}{\rule{5mm}{0mm}}

\begin{document}

\subsection*{Fracture Mechanics implications for apparent static friction
coefficient in contact problems involving slip-weakening laws}

\subsection*{A. Papangelo$^1$, M. Ciavarella$^1$, J.R.Barber$^2$}

$^1$ Dept. of Mechanics, Mathematics and Management, Politecnico di Bari, Bari - Italy, 

$^2$ Department of Mechanical Engineering,
University of Michigan, Ann Arbor, MI 48109-2125, U.S.A.

\subsection*{Abstract}

We consider the effect of differing coefficients of static and dynamic
friction coefficients on the behaviour of contacts involving microslip. The
classic solutions of Cattaneo and Mindlin are unchanged if the transition in
coefficients is abrupt, but if it occurs over some small slip distance, the
solution has some mathematical similarities with those governing the normal
tractions in adhesive contact problems. In particular, if the transition to
dynamic slip occurs over a sufficiently small area, we can identify a `JKR'
approximation, where the transition region is condensed to a line. A local
singularity in shear traction is then predicted, with a stress-intensity
factor that is proportional to the the square root of the local contact
pressure and to a certain integral of the friction coefficient-slip distance
relation. We can also define an equivalent of the `small-scale yielding'
criterion, which enables us to assess when the singular solution provides a
good approximation. One consequence of the results is that the static
coefficient of friction determined from force measurements in experiments is
significantly smaller than the value that holds at the microscale.

\section{Introduction}

If a deformable structure with frictional interfaces is subjected to loads that are insufficient to cause gross slip (sliding), the deformation of the components generally permits some local regions of `microslip' at the nominally stuck contact interfaces. When the loading is periodic, these regions contribute to the energy dissipation in the structure and hence influence the dynamic behaviour \cite{Johnson,Popp}. Also, cyclic microslip can eventually lead to the initiation and propagation of 
fretting fatigue cracks \cite{Nowell}.

Most of the extensive literature on problems involving microslip assumes
that Coulomb's friction law applies --- i.e. 
\begin{eqnarray}
\mbox{\boldmath $q$\unboldmath}&=&-fp\,\frac{\dot{%
\mbox{\boldmath
$u$\unboldmath}}}{|\dot{\mbox{\boldmath $u$\unboldmath}}|}\;;\;\;\;\dot{%
\mbox{\boldmath $u$\unboldmath}}\neq0  \label{fslip} \\
|\mbox{\boldmath $q$\unboldmath}|&\le&fp\;;\;\;\;\rule{10mm}{0mm}\dot{%
\mbox{\boldmath $u$\unboldmath}}=0\;,  \label{fstick}
\end{eqnarray}
where $\mbox{\boldmath $q$\unboldmath}$ is the frictional (tangential)
traction, $p$ is the contact pressure, $\dot{\mbox{\boldmath $u$\unboldmath}}
$ is the local microslip velocity, and $f$ is the coefficient of friction.
In particular, it is usually assumed that the same coefficient $f$ governs
both the slip and stick regions.

By contrast, dynamicists and tribologists often make a distinction between
static and dynamic friction \cite{Rabinowicz2}, so that equations (\ref%
{fslip},\ref{fstick}) are replaced by 
\begin{eqnarray}
\mbox{\boldmath $q$\unboldmath}&=&-f_d\,p\,\frac{\dot{%
\mbox{\boldmath
$u$\unboldmath}}}{|\dot{\mbox{\boldmath $u$\unboldmath}}|}\;;\;\;\;\dot{%
\mbox{\boldmath $u$\unboldmath}}\neq0  \label{fslip1} \\
|\mbox{\boldmath $q$\unboldmath}|&\le&f_s\,p\;;\;\;\;\rule{10mm}{0mm}\dot{%
\mbox{\boldmath $u$\unboldmath}}=0\;,  \label{fstick1}
\end{eqnarray}
where $f_s, f_d$ are the static and dynamic friction coefficients
respectively. In particular, if $f_s>f_d$, this friction law provides a
mechanism for `stick-slip' frictional vibrations \cite{Popp1990}.
Numerous experimental investigations have shown differences between static
and sliding friction (e.g. \cite{Rabinowicz1}). These differences are
generally small for dry metals \cite{Feynman}, but can be substantial for
earthquake fault mechanics, where ratios as high as ten between the
coefficients have been reported \cite{Poliakov}. Rice
\cite{Rice1996} characterizes such interfaces as `strong but brittle'.

A higher coefficient of static friction can to some extent be explained by
noting that the formation of adhesive bonds, which forms the basis of Bowden
and Tabor's friction theory \cite{BT}, will be enhanced by
diffusion if asperities remain in contact for some period of time. Similar
arguments can be used to justify the `rate-state' friction model \cite{Ruina,Rate-state}

In this paper, we shall examine the effect of introducing a higher
coefficient of static friction on problems involving microslip. In the
interests of simplicity, we shall restrict attention to cases where Dundurs'
parameter $\beta=0$ (\cite{KLJ} p. 110), so there is no coupling between
normal and tangential loading, and the contact pressure can be determined
without reference to the friction law. Also, we shall illustrate our ideas
in the context of the two-dimensional Hertz problem, since this is
susceptible to simple analytical solutions, but extension to other
two-dimensional cases, and to the axisymmetric Hertz problem is routine.

\section{Evolution of frictional traction distributions}

Cattaneo \cite{Cattaneo} and later Mindlin \cite{Mindlin} considered the case where two
elastic bodies are first pressed together by a normal force $P$, which is
then held constant whilst a monotonically increasing unidirectional force $%
Q_x$ is applied. The profile of the bodies was characterized by a quadratic
initial gap function $g_0(x,y)=Ax^2+By^2$, so that the normal loading phase
is defined by the classical Hertz theory. Cattaneo and Mindlin then showed
that, subject to a small approximation associated with the local slip
direction \cite{Munisamy}, the shear traction distribution
has the form 
\begin{equation}
q_x(x,y)=f\left[p(x,y)-p^*(x,y)\right]\;,  \label{C-M}
\end{equation}
where $p(x,y)$ is the contact pressure and $p^*(x,y)$ is the contact
pressure that would be developed at some smaller normal force $P^*$ given by 
\begin{equation}
P^*=P-\frac{Q_x}{f}\;.
\end{equation}
Ciavarella \cite{Ciavarella1} and J\"{a}ger \cite{Jager} have since shown that this form of
superposition is exact for any initial gap function $g_0(x)$ in the two
dimensional case, and that it is a good approximation in the general
three-dimensional case \cite{Ciavarella2}.

\section{Static and dynamic friction}

Now consider the case where $f_s>f_d$ and the loading scenario is the same
as in the Cattaneo-Mindlin problem. We assume the existence of a slip zone
in which $q_x(x,y)=f_d\,p(x,y)$, so we write the complete shear traction
distribution as 
\begin{equation}
q_x(x,y)=f_d\,p(x,y)- q_x^*(x,y)\;,  \label{corrective}
\end{equation}
where $q_x^*(x,y)$ is a corrective distribution to be
determined from the condition that the slip displacement (i.e. the relative
tangential displacement) is zero in the stick area $\mathcal{A}_{\mathrm{%
stick}}$. Conditions (\ref{fslip1},\ref{corrective}) require that $%
q_x^*(x,y) $ be non-zero only in $\mathcal{A}_{\mathrm{stick}}$, and hence
the stick condition defines a well-posed boundary-value problem for $%
q_x^*(x,y)$. The inequality condition (\ref{fstick1}) precludes
singularities in the shear tractions, and this imposes uniqueness on the
solution for any given $Q_x<f_d\,P$. It is clear that the original
Cattaneo-Mindlin solution (\ref{C-M}) with $f=f_d$ satisfies these
conditions, including the inequality, since in $\mathcal{A}_{\mathrm{stick}}$%
, this would give $q_x(x,y)<f_d\, p(x,y)<f_s\,p(x,y)$.

\section{Dependence on slip distance}

The discussion so far is predicated on the assumption that as soon as stick
is `broken' there is an immediate transition to the dynamic coefficient $f_d$%
, but in practice we might expect a more continuous transition as slip
occurs. We shall therefore examine the consequences of a friction law in
which the coefficient of friction is a continuous and monotonic function $%
f(u)$ of the slip displacement $u$, such that 
\begin{equation}
f(0)=f_s\rule{5mm}{0mm}\mbox{and}\ f(u)\rightarrow
f_d\;;\;\;\;u\rightarrow\infty\;.  \label{f(u)}
\end{equation}
Such a law can be regarded as a special case of the rate-state law \cite{Ruina,Rate-state} and is also related to the the shear
failure law proposed by Abercrombie and Rice \cite{Abercrombie}. Applications of similar
laws to fault mechanics are discussed by Ben Zion \cite{Ben-Zion}.

In general, solutions of the corresponding contact problem will then require
numerical solution, but it is instructive to consider some simple cases
analytically. In particular, we shall consider the two-dimensional case
where the bodies comprise a cylinder of radius $R$ and a half space, so the
contact pressure is given by 
\begin{equation}
p(x)=\frac{E^{\raisebox{0.7mm}{$ *$}}\sqrt{a^2-x^2}}{2R}\;;\;\;\;P=\frac{\pi E^{\raisebox{0.7mm}{$ *$}} a^2}{4R}\;,
\end{equation}
where $a$ is the semi-width of the contact area $-a<x<a$, and $E^{%
\raisebox{0.7mm}{$ *$}}$ is the composite elastic modulus \cite{KLJ}.

We anticipate the existence of two symmetric slip regions $-a<x<-c$ and $%
c<x<a$ in which the slip displacement increases monotonically away from the
stick-slip boundaries $x=\pm c$. Two limiting cases can also be identified.
If $f(u)$ is a rather slowly decaying function of $u$, the friction
coefficient will be close to $f_s$ throughout the slip regions and the
solution will approximate the constant coefficient case with $f=f_s$. At the
other limit, if a very small amount of slip displacement is required to
precipitate the change in coefficient, most of the slip area will be at or
near $f_d$, but we must still allow for the existence of `transition'
regions $c<|x|<b$ in which $f>f_d$.

The exact form of the function $f(u)$ is not critical, but it is convenient
to define a quantity $W$ with the dimensions of surface energy through the
relation 
\begin{equation}
W=\int_0^\infty (f(u)-f_d)p\,du\;,  \label{Maugis}
\end{equation}
which is equivalent to the shear fracture energy defined by Abercrombie and
Rice \cite{Abercrombie}. The contact pressure $p$ will generally vary in the transition
region, but if this is sufficiently short for $p$ to be regarded as uniform,
we can also define a length scale $\Delta$ characterizing the amount of slip
needed to transition to dynamic friction, such that 
\begin{equation}
\Delta=\frac{W}{(f_s-f_d)p}=\frac{1}{(f_s-f_d)}\int_0^\infty
(f(u)-f_d)\,du\;.  \label{Maugis10}
\end{equation}
Rabinowicz \cite{Rabinowicz1} conducted some simple but elegant experiments to determine 
$f_s,f_d$ and $\Delta$ for metals, his results\footnote{%
It is difficult to explain why different results might be obtained by simply
interchanging the materials in the mild steel/copper case, but the
difference is arguable within the range of likely experimental variance.}
being presented in Table 1.

\begin{center}
\begin{tabular}{|l|r|r|c|}
\hline
Materials & $f_s$ & $f_d$ & $\Delta$ ($\mu$m) \\ \hline
copper/mild steel & 0.46 & 0.31 & 1 \\ 
lead/mild steel & 0.72 & 0.47 & 3 \\ 
mild steel/copper & 0.54 & 0.39 & 0.9 \\ 
mild steel/titanium & 0.63 & 0.45 & 6 \\ 
mild steel/zinc & 0.65 & 0.47 & 2 \\ \hline
\end{tabular}
\end{center}

\noindent \textit{Table 1:} Friction coefficients and slip length $\Delta$
for some metal combinations, from \cite{Rabinowicz1}

\vspace{2mm} A special case satisfying equations (\ref{Maugis}, \ref%
{Maugis10}) is the step function $f=f_s-(f_s-f_d)H(u-\Delta)$, where $%
H(\cdot)$ is the Heaviside step function. The perceptive reader will notice
a similarity here to Maugis' approximate formulation of the normal adhesive
contact problem \cite{Maugis}, where the adhesion law is also represented
by a step function and the outer boundary of the adhered region is
determined from the condition that the separation there is equal to a
critical value. Indeed we shall see that there are significant mathematical
analogies between the present problem and adhesive problems.

\subsection{A double-Cattaneo-Mindlin solution}

The present problem could be formulated using a step function for $f(u)$,
but a simpler mathematical approximation can be obtained by adapting the
`double-Hertz' concept of Greenwood and Johnson \cite{Greenwood}. We first note that
the Cattaneo-Mindlin traction distribution $q_x(x)=q(x,a,c), Q_x=Q(a,c)$,
where 
\begin{equation}
q(x,a,c)=\sqrt{a^2-x^2}-\sqrt{c^2-x^2}\;;\;\;\;Q(a,c)=\frac{\pi(a^2-c^2)}{2}
\label{CM-q}
\end{equation}
produces slip displacements $u_x(x)$, such that 
\begin{eqnarray}
\frac{\partial u_x}{\partial x}\equiv v(x,a,c)&=&0\;;\rule{25mm}{0mm}-c<x<c
\label{CM-v1} \\
&=&-\frac{2\sqrt{x^2-c^2}}{E^{\raisebox{0.7mm}{$ *$}}}\;\;\;\;\;\;;\;c<|x|<a
\label{CM-v2}
\end{eqnarray}
(\cite{KLJ} p. 214), where the square roots in (\ref{CM-q}) are to be
interpreted as zero in any region where their respective arguments are
negative.

We next approximate the solution to the frictional problem as 
\begin{equation}
q_x(x)=\frac{E^{\raisebox{0.7mm}{$ *$}} f_d\,q(x,a,c)}{2R} +
Cq(x,b,c)\;;\;\;\;Q_x=\frac{E^{\raisebox{0.7mm}{$ *$}} f_d\,Q(a,c)}{2R} +
CQ(b,c)\;,  \label{dH}
\end{equation}
where $c<b<a$. The corresponding slip displacements will then satisfy 
\begin{equation}
\frac{\partial u_x}{\partial x}(x)=\frac{E^{\raisebox{0.7mm}{$ *$}}
f_d\,v(x,a,c)}{2R} + Cv(x,b,c)\;,
\end{equation}
and this is zero in $-c<x<c$ from (\ref{CM-v1}), showing that the stick
condition can be satisfied by an appropriate rigid-body translation.

The shear tractions in $b<|x|<a$ are 
\begin{equation}
q_x(x)=\frac{E^{\raisebox{0.7mm}{$ *$}} f_d\,\sqrt{a^2-x^2}}{2R}=f_d\,p(x)\;,
\end{equation}
and hence satisfy the slip condition at $f=f_d$, since the other square-root
terms make no contribution in this range. In $c<|x|<b$, the shear tractions
are 
\begin{equation}
q_x(x)=f_d\,p(x)+C\sqrt{b^2-x^2}\;,  \label{qx}
\end{equation}
and we can choose the constant $C$ so as to ensure that $q_x(c)=f_s\,p(c)$,
giving 
\begin{equation}
C=\frac{E^{\raisebox{0.7mm}{$ *$}}(f_s-f_d)}{2R}\sqrt{\frac{a^2-c^2}{b^2-c^2}%
}  \label{C}
\end{equation}
and 
\begin{equation}
q_x(x)=\frac{E^{\raisebox{0.7mm}{$ *$}} f_d\,q(x,a,c)}{2R} + \frac{E^{%
\raisebox{0.7mm}{$ *$}}(f_s-f_d)}{2R}\sqrt{\frac{a^2-c^2}{b^2-c^2}}%
\,q(x,b,c)\;.  \label{qx2CM}
\end{equation}
With this choice, the effective local coefficient of friction $f=q_x/p$ will
decrease monotonically from $f_s$ to $f_d$ in $c<|x|<b$.

The final step is to determine the unknown radii $c,b$ from the equilibrium
condition (\ref{dH})$_2$, and from (\ref{Maugis}) which we can write as 
\begin{equation}
W=\int_c^b\left[q_x(x)-f_d\,p(x)\right]\frac{du_x}{dx}dx\;.  \label{Maugis1}
\end{equation}
In $c<|x|<b$, we have 
\begin{equation}
\frac{du_x}{dx}=-\frac{1}{R}\left(f_d+(f_s-f_d)\sqrt{\frac{a^2-c^2}{b^2-c^2}}%
\right)\sqrt{x^2-c^2}\;,
\end{equation}
from (\ref{CM-v2},\ref{C}). Using this expression and (\ref{qx}) in (\ref%
{Maugis1}) and evaluating the integral, we obtain 
\begin{eqnarray}
W&=&-\frac{E^{\raisebox{0.7mm}{$ *$}} b(f_s-f_d)}{6R^2}\sqrt{\frac{a^2-c^2}{%
b^2-c^2}}\left(f_d+(f_s-f_d)\sqrt{\frac{a^2-c^2}{b^2-c^2}}\,\right)  \notag
\\
&&\rule{5mm}{0mm}\times\left[(b^2+c^2)E(k)-2c^2K(k)\right]\;,  \label{W1}
\end{eqnarray}
where 
\begin{equation}
k^2=1-\frac{c^2}{b^2}
\end{equation}
and 
\begin{equation}
K(k)=\int_0^{\pi/2}\frac{d\theta}{\sqrt{1-k^2\sin^2\theta}}%
\;;\;\;\;E(k)=\int_0^{\pi/2}\sqrt{1-k^2\sin^2\theta}\,d\theta
\end{equation}%
are the complete elliptic integrals of the first and second kind
respectively. The equilibrium condition is obtained from (\ref{CM-q},\ref{dH}%
,\ref{C}) as 
\begin{equation}
Q_x=\frac{\pi E^{\raisebox{0.7mm}{$ *$}}}{4R}\left[f_d(a^2-c^2)+(f_s-f_d)%
\sqrt{(b^2-c^2)(a^2-c^2)}\right]\;.  \label{Qx}
\end{equation}
If $Q_x, W$ are given, (\ref{W1}, \ref{Qx}) provide two equations for the
two unknown radii $c,b$.

\subsection{The `JKR' limit}

If the transition from $f_s$ to $f_d$ occurs over a sufficiently small
region, we can obtain a limiting solution analogous to the JKR solution of
normal adhesion problems. We write $b=c+\delta$, where $\delta\ll c$, in
which case (\ref{W1}) can be approximated as 
\begin{equation}
W\approx\frac{\pi E^{\raisebox{0.7mm}{$ *$}}(f_s-f_d)^2(a^2-c^2)\delta}{16R^2}%
\rule{5mm}{0mm}\mbox{implying}\rule{5mm}{0mm}\delta \approx\frac{16R^2W}{\pi E%
^{\raisebox{0.7mm}{$ *$}}(f_s-f_d)^2(a^2-c^2)}\;.  \label{delta}
\end{equation}
Also, the second term in $q_x(x)$ in equation (\ref{dH}) can be approximated
as 
\begin{equation}
Cq(x,b,c)\approx Cq(x,c,c)+C\delta\frac{\partial q}{\partial a}(x,c,c)=\frac{%
Cc\delta}{\sqrt{c^2-x^2}}\;.  \label{30}
\end{equation}
Applying the same approximation to equations (\ref{C}, \ref{Qx}) and
substituting for $\delta$ from (\ref{delta}), we obtain 
\begin{equation}
q_x(x)\approx\frac{E^{\raisebox{0.7mm}{$ *$}} f_d\,q(x,a,c)}{2R}+\sqrt{\frac{%
2WE^{\raisebox{0.7mm}{$ *$}} c}{\pi(c^2-x^2)}}\;,  \label{sing}
\end{equation}
and 
\begin{equation}
\frac{Q_x}{f_dP}=\frac{4RQ_x}{\pi E^{\raisebox{0.7mm}{$ *$}} f_da^2}\approx1-%
\frac{c^2}{a^2}+\frac{4R}{f_d\,a^2}\sqrt{\frac{2Wc}{\pi E^{%
\raisebox{0.7mm}{$
*$}}}}\;.  \label{QJKRW}
\end{equation}

Equation (\ref{sing}) defines a locally singular field, implying the
existence of a mode II stress-intensity factor 
\begin{equation}
K_{\mathrm{II}}\equiv\lim_{x\rightarrow c^-}q_x(x)\sqrt{2\pi(c-x)}=\sqrt{%
2WE^{\raisebox{0.7mm}{$ *$}}}\;,  \label{KII}
\end{equation}
which is exactly analogous with the mode I stress intensity factor $K_{%
\mathrm{I}}=\sqrt{2\Delta\gamma E^{\raisebox{0.7mm}{$ *$}}}$ in normal
adhesion problems in the JKR limit, where $\Delta\gamma$ is the interface
energy.

In an impressive series of experiments, Svetlizky and Fineberg \cite{Fineberg} have
observed frictional slip progressing by the relatively slow propagation of
slip zones behind which the shear tractions approximate a square-root
singularity. The strength of this singularity is approximately constant,
indicating a well-defined value of fracture energy $W$, but they suggest it
may depend on the local pressure, as a result of the area of actual contact
being approximately proportional to pressure.

Ciavarella \cite{Ciavarella3} presented solutions of contact problems with a mode II
stress-intensity factor around the stick-slip boundary, motivated by
Fineberg's observations. The present analysis shows that such an effect can
be generated by a slip-dependent friction law of the form (\ref{f(u)}) and
provides a rationale for determining an appropriate value of $K_{\mathrm{II}%
} $. In particular, we notice from (\ref{KII}) that the stress-intensity
factor depends only on the composite modulus and $W$, and is otherwise
independent of the details of the contact problem. Since \textit{ex hypothesi%
}, the transition is assumed to occur over a small region (of width $\delta$%
) in the contact area, we can assume that the contact pressure $p$ is
uniform in this region, and hence use the form (\ref{Maugis10}) for $W$.
This leads to a stress-intensity factor 
\begin{equation}
K_{\mathrm{II}}=\sqrt{2E^{\raisebox{0.7mm}{$ *$}}(f_s-f_d) p\Delta}\;,
\label{KIIa}
\end{equation}
which varies with $\sqrt{p}$ and is equivalent to the `pressure-dependent
toughness' criterion of \cite{Ciavarella3}.

Using (\ref{Maugis10}) to recast equations (\ref{sing}, \ref{QJKRW}) in
terms of $\Delta $, we have 
\begin{eqnarray}
q_{x}(x) &\approx &\frac{E^{\raisebox{0.7mm}{$ *$}}f_{d}\,q(x,a,c)}{2R}+E^{%
\raisebox{0.7mm}{$ *$}}\sqrt{\frac{(f_{s}-f_{d})\Delta c\sqrt{a^{2}-c^{2}}}{%
\pi R(c^{2}-x^{2})}}  \label{qxJKRDelta} \\
\frac{Q_{x}}{f_{d}P} &\approx &1-\frac{c^{2}}{a^{2}}+\frac{4}{f_{d}\,a^{2}}%
\sqrt{\frac{(f_{s}-f_{d})R\Delta c\sqrt{a^{2}-c^{2}}}{\pi }}\;.  \label{QJKR}
\end{eqnarray}

\subsection{Small-scale transition zone}

Equation (\ref{KIIa}) implies that at a sufficiently small distance $s$ from
the stick boundary, the frictional tractions have the singular form 
\begin{equation}
q_x(s)\approx f_dp+\sqrt{\frac{E^{\raisebox{0.7mm}{$ *$}}(f_s-f_d)p\Delta}{%
\pi s}}\;.
\end{equation}
However, this expression violates the stick condition (\ref{fstick1}) in the
region $0<s<s_0$, where 
\begin{equation}
\sqrt{\frac{E^{\raisebox{0.7mm}{$ *$}}(f_s-f_d)p\Delta}{\pi s_0}}=(f_s-f_d)p 
\rule{5mm}{0mm}\mbox{or}\ssa s_0=\frac{E^{\raisebox{0.7mm}{$ *$}}\Delta}{%
\pi(f_s-f_d)p}\;.  \label{s0}
\end{equation}
An analogous situation is encountered in elastic-plastic fracture mechanics,
where the `small-scale yielding' criterion is used to determine whether the
fields far outside the yield zone can reasonably be described by the elastic
solution \cite{Rice1974}. In the present case, the singular solution can be
expected to give good results everywhere except very close to $x=c$,
provided $s_0\ll c$.

This criterion depends on $c$ and hence on $Q_x$, but a rough estimate of
the applicability of the JKR solution in the present problem can be obtained
by using $p(0),a$ for $p,c$ respectively, defining the modified criterion 
\begin{equation}
\Lambda\equiv\frac{R\Delta}{(f_s-f_d)a^2}\ll1\;.  \label{Lambda}
\end{equation}

\subsection{More general two-dimensional problems}

We have analyzed the two-dimensional Hertzian problem in detail because the
resulting expressions are algebraically straightforward, enabling the
fundamental structure of the solution to be exposed. However, the same
method can be applied to any two-dimensional problem involving a single
symmetric contact area. We simply replace equation (\ref{CM-q}) by 
\begin{equation}
q(x,a,c)=p(x,a)-p(x,c)\;;\;\;\;Q(a,c)=P(a)-P(c)\;,
\end{equation}
where $p(x,a)$ is the normal contact pressure when the contact area is
defined by $-a<x<a$, and $P(a)$ is the corresponding normal force. We know
from Ciavarella \cite{Ciavarella1} and J\"{a}ger \cite{Jager} that this will satisfy equation
(\ref{CM-v1}), so the traction distribution 
\begin{equation}
q_x(x)=f_d q(x,a,c)-Cq(x,b,c)
\end{equation}
will satisfy the stick conditions in $-c<x<c$ and the dynamic slip
conditions in $b<|x|<a$. The rest of the solution can then be completed as
in \S 4.1.

If the length scale $s_0$ in (\ref{s0}) is sufficiently small to justify the
JKR approximation, the second term will take the universal form (\ref{30}),
so the solution can be written down as the superposition of a conventional
Cattaneo-Mindlin solution with coefficient of friction $f_d$ and equation (%
\ref{30}). In this context, it may be helpful to note that the limiting
expression for $Q(b,c)$ is 
\begin{equation}
Q(b,c)=\int_{-c}^c Cq(x,b,c)dx\rightarrow\sqrt{2\pi WE^{%
\raisebox{0.7mm}{$
*$}} c}\;,
\end{equation}
so the total tangential force is 
\begin{equation}
Q_x=f_d\left[P(a)-P(c)\right]+\sqrt{2\pi WE^{\raisebox{0.7mm}{$ *$}} c}\;.
\end{equation}
Since $Q_x$ will usually be prescribed, this provides an equation from which 
$c$ can be determined as a function of $Q_x$.

\section{Finite element results}

The double Cattaneo-Mindlin solution is approximate in the sense that we are
able to match a specific value of the fracture energy $W$ or (equivalently)
the length scale $\Delta$, but the exact form of the function $f(u)$ cannot
be prescribed. The implied form of this function depends on the
dimensionless ratios $b/c, a/c$, some representative curves being shown as
dashed and dotted lines in Fig. 1.


\begin{center}
\includegraphics[height=57mm]{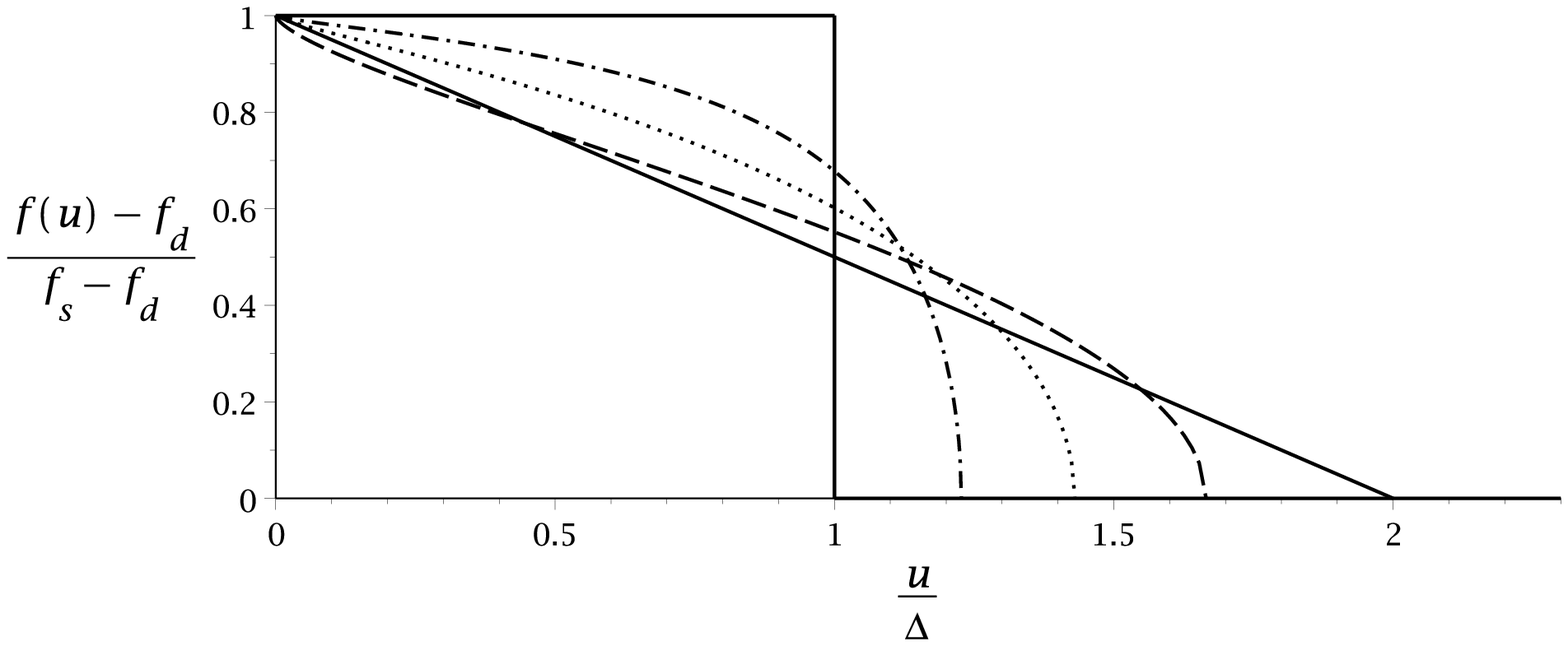}
\end{center}

\noindent \textit{Fig. 1:} The friction coefficient function $f(u)$ implied
by the double Cattaneo-Mindlin solution for $a/c=1.6, b/c=1.5$ (%
\rule[1mm]{2mm}{0.4mm} \rule[1mm]{0.4mm}{0.4mm} \rule[1mm]{2mm}{0.4mm} 
\rule[1mm]{0.4mm}{0.4mm} \rule[1mm]{2mm}{0.4mm}), $a/c=8.0, b/c=4.5$ (%
\rule[1mm]{0.5mm}{0.5mm} \rule[1mm]{0.5mm}{0.5mm} \rule[1mm]{0.5mm}{0.5mm} 
\rule[1mm]{0.5mm}{0.5mm}), $a/c=8.0, b/c=1.2$ (\rule[1mm]{2mm}{0.4mm} 
\rule[1mm]{2mm}{0.4mm} \rule[1mm]{2mm}{0.4mm}).

\vspace{2mm}

To assess the effect of this approximation, we constructed a finite element
solution of the problem, as an extension of the "verification manual" VM272
example in Ansys 15 \cite{ANSYS}, which in turn is based on the method of Yang \textit{et
al.} \cite{FEM} and an example given therein which compares satisfactorily with
the analytical Cattaneo-Mindlin solution. It is based on a mortar
formulation of the contact which is able to deal with nonconforming
discretizations across boundaries and large sliding which is more than
adequate for our problem. In \cite{FEM}, several examples
and comparisons are made to show that this method has an optimal convergence
rate and robustness with respect to other approaches. The example considers
two parallel linear elastic half cylinders of radius $R$ and pressed by a
small distributed pressure on the diameter. A tangential pressure is then
applied to cause friction at the contact interface, while the top of the
upper cylinder is constrained from rotating. The bottom of the lower
cylinder is fixed in all directions. The standard input listing available in
ANSYS is adequate for many problems, but two minor changes made in the
present case were:-

\begin{enumerate}
\item We used quadratic PLANE183-CONTA172 instead of linear elements
PLANE182-CONTA171, and we modified the mesh parametrically keeping the same
ratio of elements, in order to improve marginally the accuracy of the
results. For the figures reported in the paper we divided every element edge
by 3 which brings the total number of elements to about 45000, but still
permits a solution of an entire curve of loading in less than a minute.

\item We did not use the ANSYS variant of the friction law with just static
and dynamic coefficients, since this does not permit a dependence on slip
displacement. Instead., we defined a table of friction coefficients in terms
of slip displacement.
\end{enumerate}

Fig. 2 compares the shear traction $q_x(x)$ from equation (\ref{qx2CM}) for $%
Q_x=0.8f_dP,\, f_s=0.15, f_d=0.1, \Lambda=0.05$, with finite element results
using the ramp (linear) function for $f(u)$ from Fig. 1. The agreement is
clearly extremely good. Also shown on this figure are the conventional
Cattaneo-Mindlin prediction (equivalent to taking $\Delta=0$) and the JKR
approximation (\ref{qxJKRDelta}). The latter gives good predictions
everywhere except in the transition region, where of course the predicted
singular stress is unphysical.


\begin{center}
\includegraphics[height=55mm]{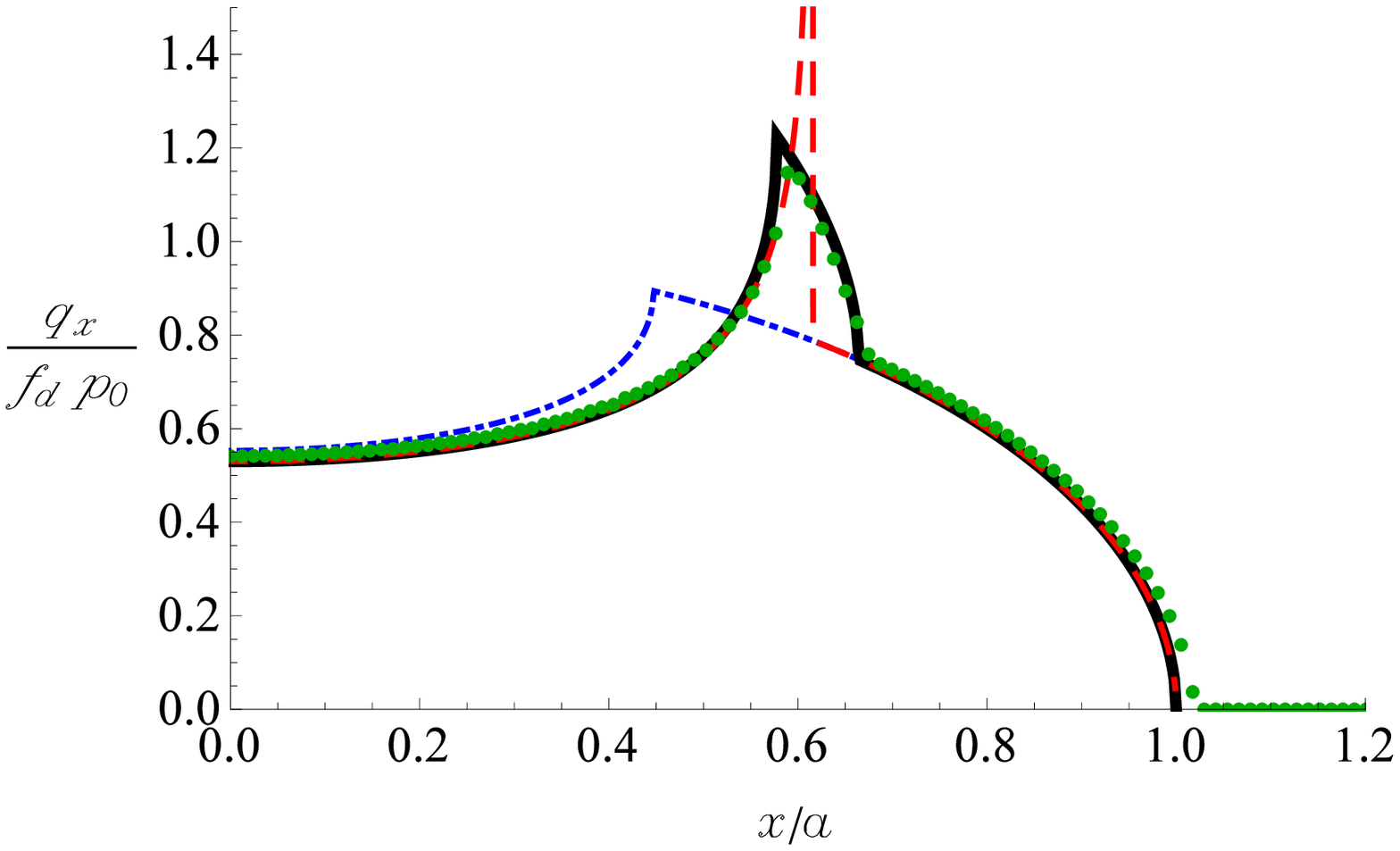}
\end{center}

\noindent \textit{Fig. 2:} Finite element results ({\color{green} $\bullet
\bullet \bullet$}) for the shear traction distribution $q_x(x)$ for $%
Q_x=0.8f_dP, f_s=0.15, f_d=0.1$ and $\Lambda=0.05$: \rule[1mm]{6mm}{0.7mm}
Double Cattaneo-Mindlin solution (\ref{qx2CM}), {\color{blue} 
\rule[1mm]{2mm}{0.4mm} \rule[1mm]{0.4mm}{0.4mm} \rule[1mm]{2mm}{0.4mm} 
\rule[1mm]{0.4mm}{0.4mm} \rule[1mm]{2mm}{0.4mm}} Conventional
Cattaneo-Mindlin solution (\ref{C-M}) with $f=f_d$, {\color{red} 
\rule[1mm]{2mm}{0.4mm} \rule[1mm]{2mm}{0.4mm} \rule[1mm]{2mm}{0.4mm} } `JKR'
approximation of equation (\ref{qxJKRDelta}).

\vspace{2mm} Fig. 3 shows a similar comparison for a larger value of $\Delta$%
, so that the transition extends over a larger radius. In this figure, we
compare equation (\ref{qx2CM}) with finite element solutions using the ramp
function and the step function respectively from Fig. 1. This figure shows
that the traction distribution is relatively insensitive the the form of the
function $f(u)$ for given values of $(f_s-f_d)$ and $\Delta$, and hence that
equation (\ref{qx2CM}) can be expected to give good results for most
practical slip-weakening laws.


\begin{center}
\includegraphics[height=55mm]{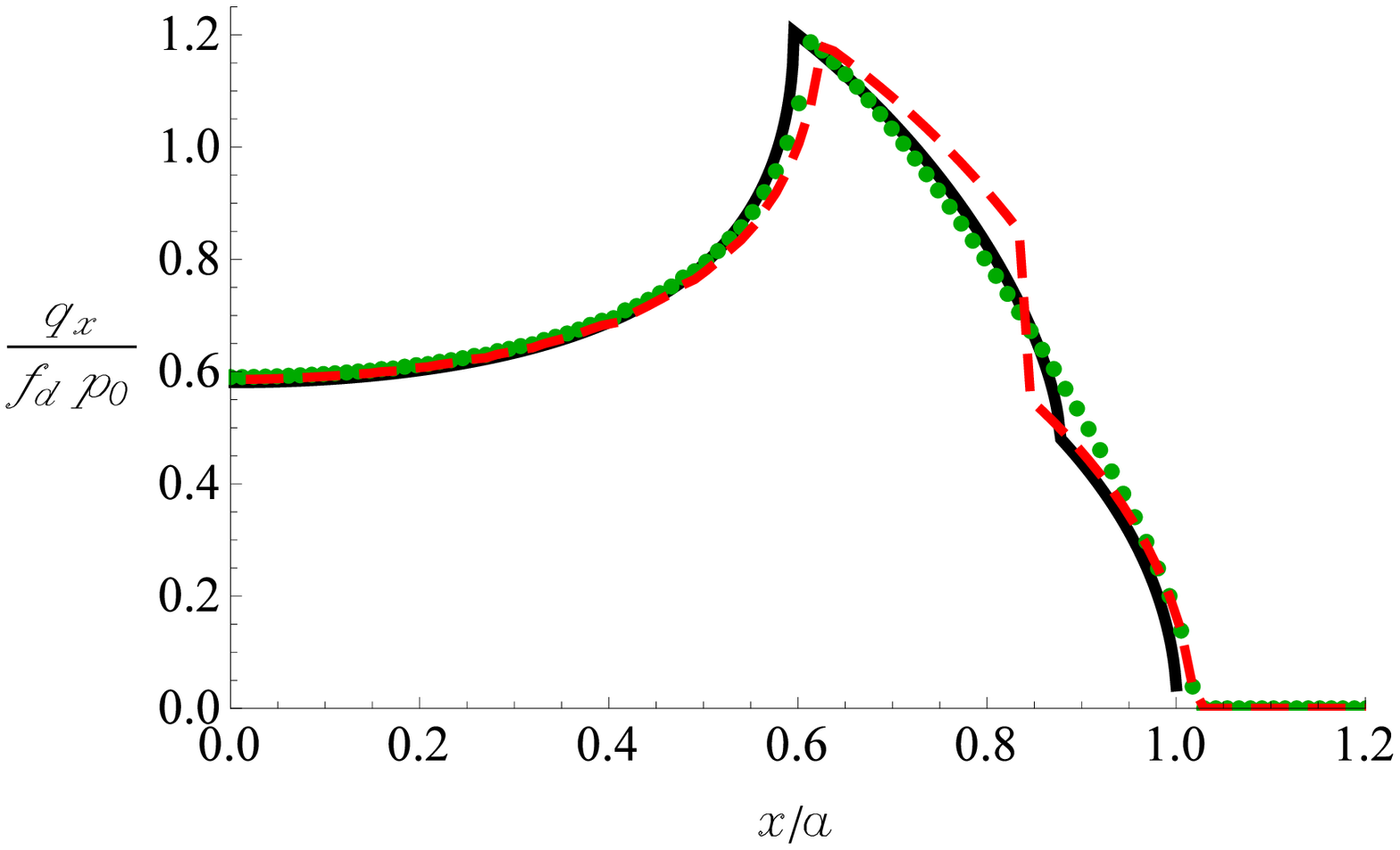}
\end{center}

\noindent \textit{Fig. 3:} Effect of the function $f(u)$ on the traction
distribution $q_x(x)$: {\color{green} $\bullet \bullet \bullet$} ramp
(finite element), {\color{red} 
\mbox{\rule[1mm]{2mm}{0.4mm}
\rule[1mm]{2mm}{0.4mm} \rule[1mm]{2mm}{0.4mm}}} step (finite element), 
\rule[1mm]{6mm}{0.7mm} equation (\ref{qx2CM}). $Q_x=0.9f_dP$, $f_s=0.15,
f_d=0.1, \Lambda=0.277$.

\section{Discussion}

The principal new result from this analysis is that fracture mechanics
concepts are introduced into the microslip problem, even when the friction
law is merely an extension of the Coulomb law allowing differing static and
dynamic coefficients. In particular, if the coefficient of friction varies
with slip dispacement over a relatively short slip distance $\Delta$, we can
determine a critical stress intensity factor or fracture toughness (\ref%
{KIIa}) that depends only on the static and dynamic coefficients, the form
of the slip-weakening law, the composite modulus and the local contact
pressure.

Equation (\ref{QJKR}) defines the relation between tangential force $Q_x$
and the semi-length $c$ of the stick area in the JKR limit, which is
appropriate if the small-scale transition criterion $s_0\ll c$ is satisfied.
It is plotted in Fig. 4 for several values of the dimensionless parameter 
\begin{equation}
\Psi=\left(\frac{f_s}{f_d}-1\right)\frac{R\Delta}{f_da^2}=\left(\frac{f_s}{%
f_d}-1\right)^2\Lambda\;.
\end{equation}


\begin{center}
\includegraphics[height=60mm]{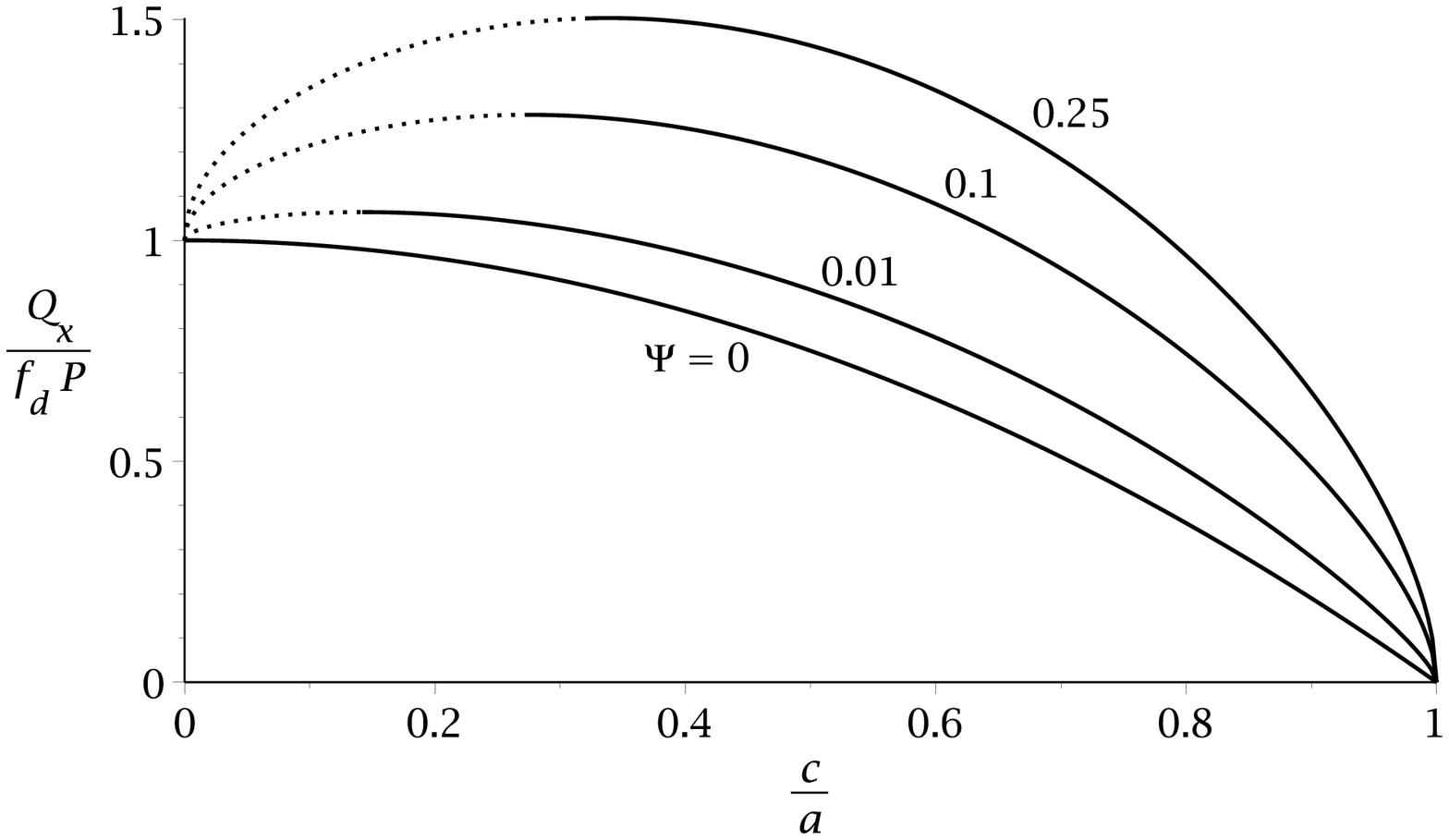}
\end{center}

\noindent \textit{Fig. 4:} The tangential force $Q_x$ as a function of the
radius $c$ of the stick zone (JKR limit).

\vspace{2mm} All the curves except the limiting case $\Psi=0$
exhibit a maximum $Q_x=Q_{\max}$ at a non-zero value of $c$, implying that
under tangential force control, the system would jump unstably to full
sliding once this maximum is reached. The unstable range is shown dotted in
Fig. 4.

Similar plots were made for the double Cattaneo-Mindlin solution, using
equations (\ref{Qx}, \ref{W1}) with $W=(f_s-f_p)\Delta$. Fig. 5
compares the resulting curves for $\psi=0.1$ and $\Lambda=0.025, 0.4$ with
the JKR solution. Notice that changing $\Lambda$ at constant $\psi$ implies
a change in the friction coefficient ratio $f_s/f_d$. The truncation in
these curves near $c=a$ occurs because the outer boundary $b$ of the
transition region cannot exceed the boundary $a$ of the contact area. When $%
b=a$, the double Cattaneo-Mindlin solution reduces to a conventional
Cattaneo-Mindlin solution with $f=f_s$, so we have arbitrarily used this
result to continue the curves to $c=a$ [shown dotted]. 


\begin{center}
\includegraphics[height=65mm]{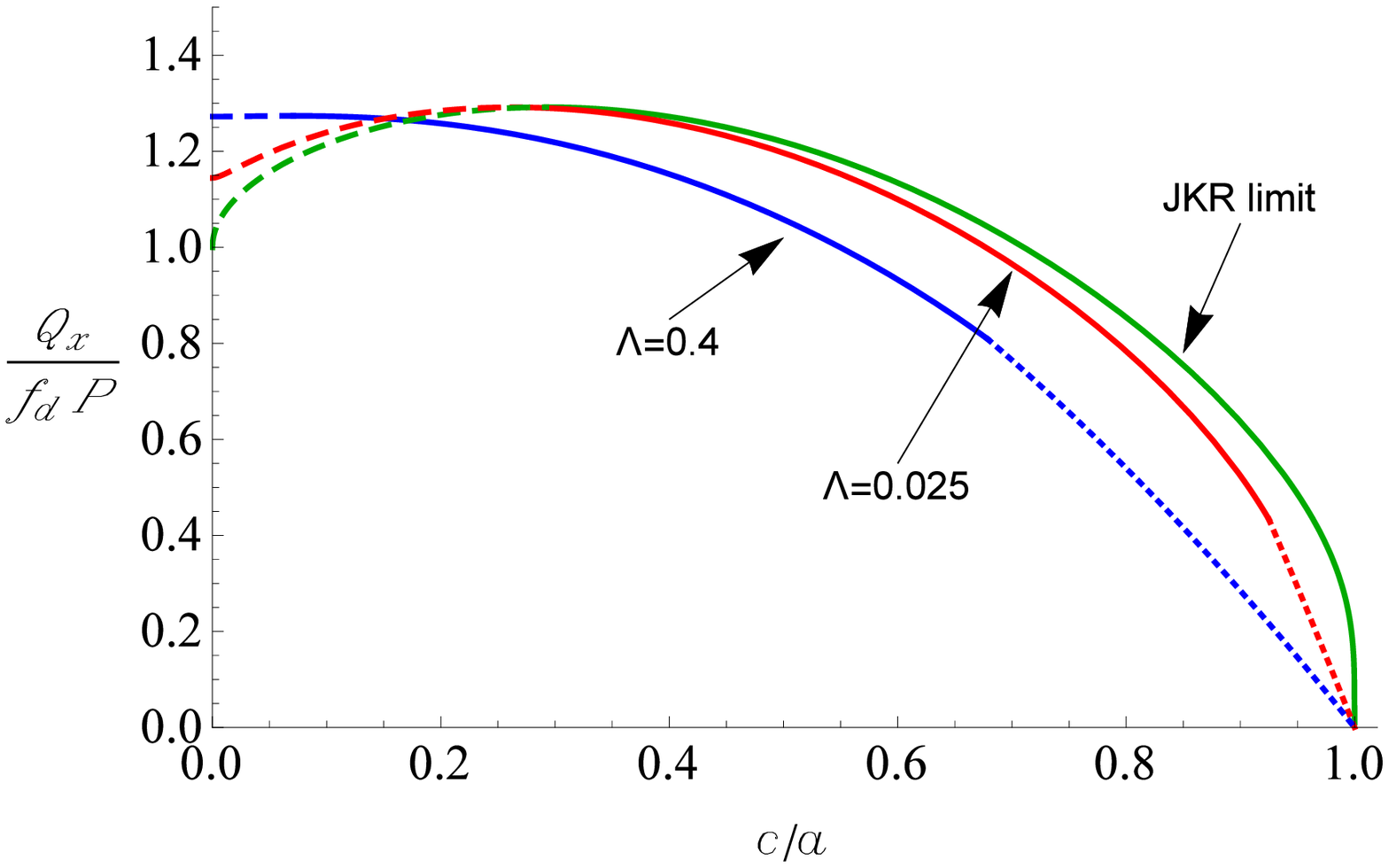}
\end{center}

\noindent {\it Fig. 5:} Comparison of the double
Cattaneo-Mindlin solution with the JKR limit for $\psi=0.1$.

As predicted, the curve for $\Lambda=0.025$ is very close to the JKR curve,
though the maximum $Q_x$ is shifted slightly to the left. Notice
incidentally that we might have chosen to plot the double Cattaneo-Mindlin
curves as functions of the location of the mid-point $(c+b)/2a$ of the
slip-stick transition region, in which case this shift would be much
reduced. For larger $\Lambda$, the maximum occurs at significantly lower
values of $c$, but $Q_{\max}$ is still very well predicted by the JKR theory
even for $\Lambda=0.4$.

Experimental measurements of static friction ceofficient are usually
obtained by increasing the applied tangential force until sliding commences.
However, it is clear that under these circumstances, microslip is likely to
occur before gross sliding commences, and hence in the present geometry such
experiments would lead to the static coefficient of friction being
identified as $Q_{\max}/P$, which generally differs from $f_s$.


\begin{center}
\includegraphics[height=65mm]{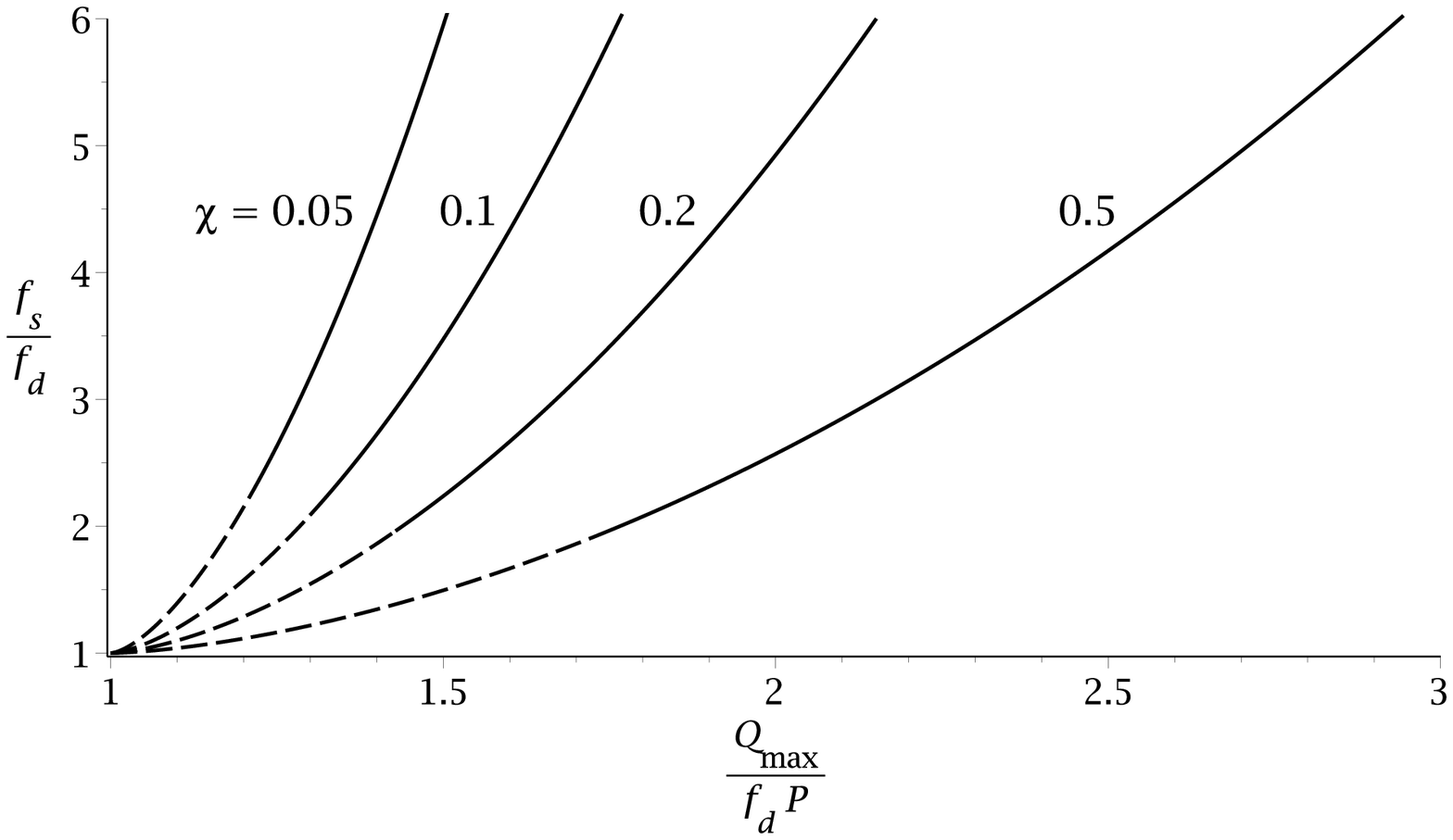}
\end{center}

\noindent {\it Fig. 6:} The coefficient ratio $f_s/f_d$ as a
function of the apparent ratio $Q_{\max}/f_d\,P$.

\vspace{2mm}  Fig. 6 shows the relationship between $f_s/f_d$
and the `apparent' value of this ratio determined as $Q_{\max}/f_d\,P$, for
various values of 
\begin{equation}
\chi=\frac{R\Delta}{f_da^2}=\left(\frac{f_s}{f_d}-1\right)\Lambda\;.
\label{chi}
\end{equation}
The dashed lines in  this figure correspond to ranges in which
the small-scale transition criterion $s_0\ll c$ is not satisfied. We notice
that the apparent static friction coefficient is always significantly lower
than $f_s$. The reason of course is that by the time $Q_{\max}$ is reached,
a significant part of the contact area has slipped sufficiently to
transition to a local coefficient $f_d$, and the measured coefficient is a
weighted average over the whole contact area.

Notice that the limiting case $\chi=0$ can arise only if $%
\Delta=0$, meaning that the transition from $f_s$ to $f_d$ occurs over an
infinitesimal slip distance. As explained in \S 3, the partial slip solution
is then identical to the conventional Cattaneo-Mindlin solution with $f=f_d$
and hence slip occurs for $Q=f_dP$ regardless of the static coefficient of
friction $f_s$. This case is defined by the vertical axis in Fig. 6.

\section{Conclusions}

We have shown that the use of a slip-weakening friction law has a
qualitative effect on the solution of microslip problems. The mechanics of
the classical Cattaneo-Mindlin problem then have a mathematical structure
similar to that of the adhesive contact problem, and we can identify an
analogue of the `JKR' solution, in which the extent of the stick zone is
governed by the occurrence of a pressure-dependent mode II stress-intensity
factor at the stick-slip boundary. By exploring the two-dimensional Hertzian
geometry, we were able to identify the equivalent fracture toughness, which
is independent of the detailed goemetry, but proportional to the square root
of the local contact pressure. We also defined a length scale $s_0$
analogous to the small-scale yielding criterion whose value enables us to
judge whether the singular solution gives a good approximation to the more
exact solution.

The tangential force reaches a maximum before the stick zone has shrunk to
zero, at which point there will be a discontinuous change of state to gross
sliding. This implies that estimates of the static coefficient of friction
from experiments on the inception of sliding will generally significantly
underestimate the values appropriate at the microscale.

\section*{Acknowledgements}

MC is grateful to the Humboldt foundation for sponsoring his visit at
Hamburg TUHH University and to J. Fineberg for stimulating discussions.


\begin{thebibliography}{30}

\bibitem{Johnson}
Johnson KL. 1961. Energy dissipation at spherical
surfaces in contact transmitting oscillating forces, \textit{J. Mech. Eng.
Sci.} \textbf{3}, 362--368.

\bibitem{Popp}
Popp K, Panning L, Sextro W. 2003. Vibration
damping by friction forces: theory and applications. \textit{J Vib Control,} 
\textbf{9}, 419--448.

\bibitem{Nowell}
Nowell D, Dini D, Hills DA. 2006. Recent
developments in the understanding of fretting fatigue. \textit{Eng Fract
Mech.} \textbf{73}, 207--222.

\bibitem{Rabinowicz2}
Rabinowicz E. 1995. \textit{Friction and Wear of
Materials}, John Wiley, New York, 2nd. edn.

\bibitem{Popp1990}
Popp K., Stelter P. 1990. Stick-slip vibrations and
chaos.\textit{Phil. Trans. R. Soc. Lond.,} \textbf{332}, 89--105.

\bibitem{Rabinowicz1}
Rabinowicz E. 1951, The nature of the static and
kinetic coefficients of friction, \textit{J. Appl. Phys.,} \textbf{22},
1373--1379.

\bibitem{Feynman}
Feynman RP, Leighton RB, Sand M. 1964. \textit{The
Feynman Lectures on Physics,} Addison-Wesley, Vol. I, 12-5.

\bibitem{Poliakov}
Poliakov ANB, Dmowska R, Rice JR. 2002. Dynamic
shear rupture interactions with fault bends and off-axis secondary faulting.%
\textit{J. Geophys. Res.,} \textbf{107 (B11)}, Art. 2295.

\bibitem{Rice1996}
Rice JR. 1996. Low-stress faulting: Strong but
brittle faults with local stress concentrations.\textit{Eos Trans. AGU,} 
\textbf{77} (46), Fall Meet. Suppl., F471.

\bibitem{BT}
Bowden FP, Tabor D. 1950. \textit{The Friction and
Lubrication of Solids,} Clarendon Press, Oxford, 1950.

\bibitem{Ruina}
Ruina, A. 1983. Slip instability and state variable
friction laws. \textit{J. Geophys. Res.,} \textbf{88 (B12)}, 10359--10370.

\bibitem{Rate-state}
Rice JR, Lapusta N., Ranjith K. 2001. Rate and
state dependent friction and the stability of sliding between elastically
deformable solids. \textit{J. Mech. Phys. Solids,} \textbf{49}, 1865--1898.

\bibitem{KLJ}
Johnson KL. 1985.\textit{Contact Mechanics},
Cambridge University Press, Cambridge.

\bibitem{Cattaneo}
Cattaneo C. 1938. Sul contatto di due corpi
elastici: distribuzione locale degli sforzi. \textit{Rendiconti
dell'Accademia Nazionale dei Lincei} \textbf{27}, 342--348, 434--436,
474--478. (In Italian)

\bibitem{Mindlin}
Mindlin RD. 1949. Compliance of elastic bodies in
contact. \textit{ASME J. Appl. Mech.} \textbf{16}, 259--268.

\bibitem{Munisamy}
Munisamy RL, Hills DA, Nowell D. 1994. Static
axisymmetrical Hertzian contacts subject to shearing forces. \textit{ASME
J.Appl. Mech.} \textbf{61}, 278--283

\bibitem{Ciavarella1}
Ciavarella M. 1998a. The generalized Cattaneo
partial slip plane contact problem. I-Theory, II-Examples.\textit{%
Int.J.Solids Struct.} \textbf{35}, 2349--2378.

\bibitem{Jager}
J\"{a}ger J. 1998. A new principle in contact mechanics. \textit{%
ASME J.Tribology.} \textbf{120}, 677--684.

\bibitem{Ciavarella2}
Ciavarella M. 1998b. Tangential loading of general
three-dimensional contacts. \textit{ASME J. Appl. Mech.} \textbf{65},
998--1003.

\bibitem{Abercrombie}
Abercrombie RE. Rice JR. 2005. Can observations of earthquake
scaling constrain slip weakening? \textit{Geophys.J. Int.,} \textbf{162},
406--424.

\bibitem{Ben-Zion}
Ben-Zion Y. 2008. Collective behavior of
earthquakes and faults: continuum-discrete transitions, progressive
evolutionary changes, and different dynamic regimes.\textit{Rev.Geophys.} 
\textbf{46}, RG4006.

\bibitem{Maugis}
Maugis, D. 1992. Adhesion of spheres: The JKR-DMT
transition using a Dugdale model. \textit{J. Colloid Interface Sci.,} 
\textbf{150}, 243--269.

\bibitem{Greenwood}
Greenwood JA, Johnson KL. 1998, An alternative to
the Maugis model of adhesion between elastic spheres. \textit{J. Phys. D:
Applied Physics.} \textbf{31}, 3279--3290.

\bibitem{Fineberg}
Svetlizky I, Fineberg J. 2014. Classical shear
cracks drive the onset of dry frictional motion, \textit{Nature} \textbf{509}%
, 205--208.

\bibitem{Ciavarella3}
Ciavarella M. 2015. Transition from stick to slip
in Hertzian contact with ``Griffith'' friction: the Cattaneo-Mindlin problem
revisited. \textit{J. Mech. Phys. Solids,} under review.

\bibitem{Rice1974}
Rice JR. 1974, Limitations to the small scale
yielding approximation for crack tip plasticity,\textit{Journal of the
Mechanics and Physics of Solids,} Vol. \textbf{22}, 17--26.

\bibitem{ANSYS}
ANSYS Mechanical APDL Verification Manual - "VM272:
2-D and 3-D Frictional Hertz Contact", ANSYS, Inc. Release 15.0 November
2013, 785--788

\bibitem{FEM}
Yang B, Laursen TA., Meng X. 2005, Two dimensional
mortar contact methods for large deformation frictional sliding.\textit{Int.
J. Numer. Meth. Engng.,} \textbf{62} 1183--1225. doi: 10.1002/nme.1222

\end{thebibliography}
\end{document}